\documentclass[prl,amsmath,amssymb,twocolumn]{revtex4}

\usepackage{graphicx}
\usepackage{subfigure}
\usepackage{dcolumn}
\usepackage{verbatim}
\usepackage{epstopdf}

\def\dd{\textrm{d}}
\def\Tr{\textrm}
\def\Bf{\boldsymbol}
\def\etal{~\textit{et~al.}}     % etal

\def\rv{\Bf{r}}

\def\Av{\Bf{A}}

\def\nv{\Bf{\nabla}}

\def\la{\langle}
\def\ra{\rangle}

\begin{document}
\title{Pairing instability driven by macroscopically degenerate collective modes in two-dimensional rotating fermion liquids near unitarity}

\author{Predrag Nikoli\'c}
\affiliation{Department of Physics, Rice University, Houston, TX 77005}

\begin{abstract}

Fermionic superfluids can undergo phase transitions into different kinds of normal regimes, loosely characterized by whether Cooper pairs remain locally stable. If the normal phase retains strong pairing fluctuations, it behaves like a liquid of vortices, which has been observed in cuprate superconductors. We argue that analogous strongly correlated normal states exist in two-dimensional neutral fermion liquids near unitarity, where superfluid is destroyed by fast rotation. These states have non-universal properties, and if they develop as distinct thermodynamic phases they can be characterized as quantum Hall states of Cooper pairs. The formal analysis is based on a model with SP($2N$) symmetry that describes the quantum critical region in the vicinity of a broad Feshbach resonance. We explore the pairing phase diagram and demonstrate that the considered model has macroscopically degenerate bosonic modes in the normal phase, to all orders in $1/N$. It takes finite-range interactions to lift this degeneracy, making the Abrikosov flux lattice of the superfluid particularly susceptible to quantum melting.

\end{abstract}

\date{\today}

\maketitle

Recent years have seen many developments regarding strongly correlated normal phases of fermionic superconductors and superfluids. It is becoming increasingly clear that the unconventional normal phase of cuprate superconductors results from quantum fluctuations that destroy the long-range phase coherence of Cooper pairs \cite{Emery}. These fluctuations can be ascribed to the motion of vortices, and indeed the normal state has properties of a vortex liquid \cite{Ong1, Ong2}. A byproduct of the fluctuations and crystal lattice is the emergence of density modulations, \cite{Zlatko1, Subir1}, which are nowdays observed in great detail \cite{Fischer, Yazdani, Hudson, Davis, Valla}.

Ultra-cold atoms, being much simpler than electronic systems like cuprates, are very attractive for studies of many-body physics \cite{Bloch}. This paper explores the possibility of obtaining analogous strongly correlated normal states using neutral cold atoms in continuum. We consider two-dimensional two-component fermion gasses in which the superfluid phase is destabilized by fast rotation at low temperatures \cite{Moller, ZhaiHo, YangZhai}. We argue that the unconventional normal states are most likely to be stable near a broad Feshbach resonance, when two-body scattering length is very large. Due to the absence of long-range forces between neutral particles, a large scattering length is what provides interactions potent enough to stabilize strongly correlated normal states. Experimental progress toward achieving such states is still limitted, but the overall pace of development in the field gives many reasons for optimism \cite{Dalibard, Engels, Stock}.

Static three-dimensional cold-atom systems near a broad Feshbach resonance have striking universal properties \cite{Bloch, Kohler, unitary, rvs}. This universality is controlled by a quantum critical point at the zero density Feshbach resonance \cite{unitary}, which defines the unitarity limit. The superfluid in this regime interpolates between the Bardeen-Cooper-Schrieffer (BCS) state of Cooper pairs and Bose-Einstein condensate (BEC) of diatomic molecules, and insulating states in optical lattices interpolate between the two analogous limits of band and Mott insulators respectively \cite{optlat}.

The situation is different in two-dimensional rotating cold atom systems because some universal properties are lost. Nevertheless, the unitarity limit is well defined because it originates in the nature of interaction potential between atoms. Consider a theory of $2N$ fermion species ($\alpha\in\lbrace\uparrow\downarrow\rbrace$, $i=1\dots N$) coupled to a bosonic Cooper pair field $\Phi$ in $d$ dimensions ($\hbar=1$):
\begin{eqnarray}\label{Action1}
S & = & \int\dd\tau\dd^dr \Bigl\lbrack \psi_{i\alpha}^\dagger \left( \frac{\partial}{\partial\tau}
    - \frac{\nv^2}{2m} - \mu \right) \psi_{i\alpha} + N \Phi^\dagger \hat{\Pi}^{(0)} \Phi \nonumber \\
 && + \Phi^\dagger \psi_{i\uparrow} \psi_{i\downarrow}
    + \Phi \psi_{i\downarrow}^\dagger \psi_{i\uparrow}^\dagger \Bigr\rbrack \ .
\end{eqnarray}
This is a generalization of the ``two-channel'' model, which we recover by setting $N=1$. The operator $\hat{\Pi}^{(0)}$ is proportional to detuning $\nu$ from the Feshbach resonance (a part of Eq.~(\ref{PiReg2})). In the absence of rotation at $T=0$, $\mu=0$ (zero density) and $\nu=0$ (Feshbach resonance) this theory is at a fixed point of renormalization group (RG), and no operators other than the written ones are relevant in $d>2$ \cite{unitary}. Therefore, only the zero-range interactions matter, and finite short-range forces have negligible impact on macroscopic properties near the fixed point. Note that no particular scale characterizes interactions, so that the usual perturbation theory in interaction strength is not applicable. For that reason we generalize to large $N$ and obtain controlled approximations by perturbative expansions in powers of $1/N$.

We now introduce rotation at angular velocity $\omega$ and a potential well to make the system quasi two-dimensional:
\begin{eqnarray}\label{RotAct}
S & = & \int\dd\tau\dd^2r \Bigl\lbrack \psi_{i\alpha}^\dagger \left( \frac{\partial}{\partial\tau}
    + \frac{(-i\nv-\Av)^2}{2m} - \mu \right) \psi_{i\alpha}  \nonumber \\
 && + N \Phi^\dagger \hat{\Pi}^{(0)} \Phi + \Phi^\dagger \psi_{i\uparrow} \psi_{i\downarrow}
    + \Phi \psi_{i\downarrow}^\dagger \psi_{i\uparrow}^\dagger \Bigr\rbrack
    \ , ~~~~~~
\end{eqnarray}
where $\nv\times\Av = \hat{z} B = \hat{z} m\omega_c = \hat{z} 2m\omega$. This action describes dynamics in the rotating frame, assuming that centrifugal forces are exactly balanced by a trap. We choose this simplification in order to focus on many-body effects without being side-tracked by trap effects.

The main utility of considering the model ~(\ref{RotAct}) comes from its simplicity and striking properties that characterize the unitarity limit. We shall reveal an unusual pairing instability in this model, which involves a macroscopic number of soft modes that are degenerate to all orders in $1/N$. An implication is that all possible arrangements of vortices in the superfluid differ very little by free energy close to the pairing transition, so that small perturbations can easily mix them and melt the vortex lattice. Note that the phases of the order parameter are only algebraically correlated in the two-dimensional superfluid with a vortex lattice 
\cite{Moore, Zlatko2, MacDonald}.

The model ~(\ref{RotAct}) contains only zero-range interactions (the Hubbard-Stratonovich field $\Phi$ has no bare dispersion at the fixed point of ~(\ref{Action1})). We demonstrate using RG that finite-range interactions between fermions are perturbations to ~(\ref{RotAct}) that can shape phases and induce transitions. Physically, this comes from the fact that the bare fermion spectrum in ~(\ref{RotAct}) consists of dispersionless macroscopically degenerate Landau levels with energies $\epsilon_n = n \omega_c - \mu'$ (we measure energy with respect to the redefined chemical potential $\mu'=\mu-\omega_c/2$). The lack of dispersion reduces dimensionality in RG and makes even arbitrarily weak perturbations extremely potent in lifting the degeneracy. We set up RG by generalizing ~(\ref{RotAct}) to $d$ dimensions in the Landau level basis using the Landau gauge, and include additional allowed terms:
\begin{widetext}
\begin{eqnarray}\label{RotAct2}
S & = & \int\dd\tau\dd^{d-2}r_{\perp} \Biggl\lbrace \sum_n
        \int\frac{\dd k_x}{2\pi} \psi_{n,k_x}^{\dagger}
          \left( \frac{\partial}{\partial\tau} + n\omega_c - \frac{\nv_{\perp}^2}{2m} - \mu' \right)
        \psi_{n,k_x}^{\phantom{dagger}}
    +   N \sum_{n_1 n_2} \int\frac{\dd p_x}{2\pi}
        \Phi_{n_1,p_x}^{\dagger} \hat{\Pi}_{n_1,n_2}^{(0)} \Phi_{n_2,p_x}^{\phantom{\dagger}}
        \nonumber \\
  & + & g \sum_{n m_1 m_2} \int\frac{\dd k_x}{2\pi} \frac{\dd p_x}{2\pi}
        \Gamma^n_{m_1 m_2} \left( \frac{k_x}{\sqrt{B}} \right) \left\lbrack \Phi_{n,p_x}^{\dagger}
          \psi_{m_1,k_x+\frac{p_x}{2}}^{\phantom{\dagger}}
          \psi_{m_2,-k_x+\frac{p_x}{2}}^{\phantom{\dagger}}
        + \Tr{h.c.} \right\rbrack \nonumber \\
  & + & u_2 \sum_{m_1 \dots m_4} \int\frac{\dd k_{x1}}{2\pi} \frac{\dd k_{x2}}{2\pi} \frac{\dd q_x}{2\pi}
        \Gamma_{m_1 \dots m_4}' \left( k_{x1}, k_{x2}, q_x \right)
        \psi_{m_1,k_{x1}}^{\dagger} \psi_{m_2,k_{x2}}^{\dagger}
        \psi_{m_3,k_{x2}+q_x}^{\phantom{\dagger}} \psi_{m_4,k_{x1}-q_x}^{\phantom{\dagger}}
        \Biggr\rbrace + \cdots \ .
\end{eqnarray}
\end{widetext}
We suppress spin $\alpha$ and flavor $i$ indices for brevity. Since the bare fermion states are localized in the plane perpendicular to the axis of rotation, it is appropriate to not rescale $k_x$ in RG. At zero temperature and density, fermion self-energy vanishes so that all renormalization comes from the boson field $\Phi$. The RG flow equations can be calculated exactly to all orders of perturbation theory when $u_2$ and the remaining omitted couplings are zero, since then the renormalization of $\Phi$ involves summation of a geometric series of bare fermion bubble diagrams. Following the procedure in Ref.\cite{unitary}, we find that under rescaling $r_{\perp}'=e^{-l}r_{\perp}$, $\tau' = e^{-2l}\tau$, $\psi'=e^{(d/2-1)l}\psi$, $\Phi'=e^{(d/2-1)l}\Phi$ the exact flow equations are:
\begin{eqnarray}
\frac{\dd\mu'}{\dd l} = 2\mu' ~~ && ~~ \frac{\dd\omega_c}{\dd l} = 2\omega_c \\
\frac{\dd\nu}{\dd l} = \left( 2 - a g^2 \right) \nu ~~ && ~~
  \frac{\dd g}{\dd l} = \left( 3 - \frac{d}{2} \right) g - b N g^3 \ , \nonumber
\end{eqnarray}
where $a$ and $b$ are cut-off dependent constants whose values are not important for the present discussion, and $\hat{\Pi}^{(0)}\propto\nu$. The interacting fixed point at $\mu'=\omega_c=\nu=0$, $g^* = \sqrt{(3-d/2)/(bN)}$ defines the unitarity limit in any dimension $d$, and shows how perturbation theory can be justified for large $N$. Appart from $\mu'$, $\omega_c$ and $\nu$, various couplings $u_n$, which may include non-local short-range potentials and hence multi-particle collision terms, can be relevant at this fixed point. For $n$-particle scattering $u_n$ we find sufficient indication at the tree-level:
\begin{equation}
\frac{\dd u_n}{\dd l} = \bigl\lbrack d + (2-d) n \bigr\rbrack u_n + \mathcal{O}(u^2)
\end{equation}
irrespective of the spatial dependence of interaction potentials in the plane perpendicular to rotation axis, since coordinates do not rescale in this plane. Therefore, all $u_n$ are relevant in $d<2$, while in $d>2$ they are relevant for $n<d/(d-2)$, with an upper bound $n \sim N$ beyond the tree-level for $N \gg 1$ and $d \ge 2$.

The physical consequences of this scaling will be apparent shortly. However, we first elucidate the effects associated with unitarity by ignoring the perturbations ($u_n \to 0$) and absorbing $g$ in the definition of $\Phi$. Hence, we now focus on the simple theory ~(\ref{RotAct}) in two dimensions.

Let us start in a quantum Hall state of unpaired fermions and consider the pairing instability. This instability shows in the poles of the boson field Green's function given by the inverse fermion bubble diagram in particle-particle channel, $G_\Phi = (N\Pi)^{-1}$. The bubble diagram in Landau representation is (without $\hat{\Pi}^{(0)}$):
\begin{eqnarray}\label{Pi1}
&& \Pi_{n,n'}(p_x,i\Omega) = \sqrt{B} \sum_{m_1,m_2} \int\frac{\dd k_x}{2\pi}
  \frac{f(\varepsilon_{m_1}) - f(-\varepsilon_{m_2})}
       {-i\Omega + \varepsilon_{m_1} + \varepsilon_{m_2}} \nonumber \\
&& ~~ \times \Gamma_{m_1,m_2}^{n*}\left(\frac{k_x}{\sqrt{B}}\right)
  \Gamma_{m_1,m_2}^{n'}\left(\frac{k_x}{\sqrt{B}}\right)
  + \mathcal{O}\left(\frac{1}{N}\right) \ .
\end{eqnarray}
Here, as in ~(\ref{RotAct2}), the Landau gauge quantum numbers $n$ and $k_x$ determine fermionic and bosonic wavefunctions:
\begin{eqnarray}\label{WF}
\psi_{n,k_x}(\rv) & = & \frac{e^{ik_xx}}{\sqrt{2^n n!}}
  \frac{e^{-\frac{B}{2}\left(y+\frac{k_x}{B}\right)^2}}{(B/\pi)^{-1/4}}
  H_n\left(\sqrt{B}y+\frac{k_x}{\sqrt{B}}\right) \\
\Phi_{n,p_x}(\rv) & = & \frac{e^{ip_xx}}{\sqrt{2^n n!}}
  \frac{e^{-B\left(y+\frac{p_x}{2B}\right)^2}}{(2B/\pi)^{-1/4}}
  H_n\left(\sqrt{2B}y+\frac{p_x}{\sqrt{2B}}\right) \ , \nonumber
\end{eqnarray}
where $H_n(x)$ are Hermite polynomials. Note that boson ``charge'' is twice that of fermions. The vertex function in Landau representation is readily derived from ~(\ref{WF}):
\begin{eqnarray}\label{Vertex1}
&& \Gamma_{m_1,m_2}^n(\xi_x) = \frac{2^{-(n+m_1+m_2)/2}}{\sqrt{\pi n!m_1!m_2!}}
  \left(\frac{2}{\pi}\right)^{\frac{1}{4}} e^{-\xi_x^2} \times \\
&&  \times \int_{-\infty}^{\infty} \dd\xi e^{-2\xi^2} H_n(\sqrt{2}\xi) H_{m_1}(\xi+\xi_x) H_{m_2}(\xi-\xi_x)
  \nonumber \ .
\end{eqnarray}
Note that the vertex does not depend on the boson momentum $p_x$, as a curious consequence of the Landau level degeneracy. The bare fermion Green's functions in Landau representation is simply $G_n(i\omega) = (-i\omega+\epsilon_n)^{-1}$.

Pairing instability can be directly found from the bubble diagram matrix $\hat{\Pi}$ at zero Matsubara frequency as the onset of a first negative eigenvalue. The matrix elements of $\hat{\Pi}$ in Landau representation are given by ~(\ref{Pi1}). We note that ~(\ref{Pi1}) is not gauge invariant, but the spectrum of $\hat{\Pi}$ is: gauge transformations are manifestly unitary transformations of $\hat{\Pi}$ in position representation. Therefore, we can safely fix gauge and look for the instability in any convenient representation.

Before proceding with calculations, it is necessary to regularize the ultra-violet divergent expression ~(\ref{Pi1}). In the absence of rotation, regularization leaves behind a chemical potential term for the $\Phi$ field, proportional to the detuning $\nu$ from the Feshbach resonance \cite{unitary}. The same procedure is problematic in the rotating case because Landau orbitals are localized and scattering length $a=-1/\nu$, whose formal definition rests on the scattering of free particles, is not entirely meaningful. Only for $a \ll R_c$, the cyclotron radius at Fermi energy, $\nu$ retains its physical meaning. We formally extend the definition of detuning from this limit to all values $\nu$. Using the relationship between $a$ and two-body scattering vertex, we regularize $\hat{\Pi}$ by adding ``$\Delta\hat{\Pi}\equiv 0$'' to ~(\ref{Pi1}):
\begin{equation}\label{PiReg2}
\Delta\Pi_{n,n'} = \left( \frac{m\nu}{4\pi a_z \Pi(0)} - 1 \right) \Pi_{n,n'}(0,0)\Bigl\vert_{\mu=T=0} \ ,
\end{equation}
where $a_z$ is a quantum well confinement length scale in $z$-direction and $\Pi(0) = \la p_\mu=0 \vert \hat{\Pi} \vert p_\mu=0 \ra$ is the projection of $\hat{\Pi}$ to plane-waves at zero momentum:
\begin{equation}\label{Pi0}
\Pi(0) = \sum_{n,n'} \Pi_{2n,2n'} \frac{(2n-1)!!(2n'-1)!!}{\sqrt{(2n)!(2n')!}} \ .
\end{equation}
The expression ~(\ref{PiReg2}) guaranties that the ultra-violet divergence is cancelled in all matrix elements, without introducing arbitrary (unknown) features at cut-off scales.

\begin{figure}[!]
\subfigure[{}]{\includegraphics[height=1.4in]{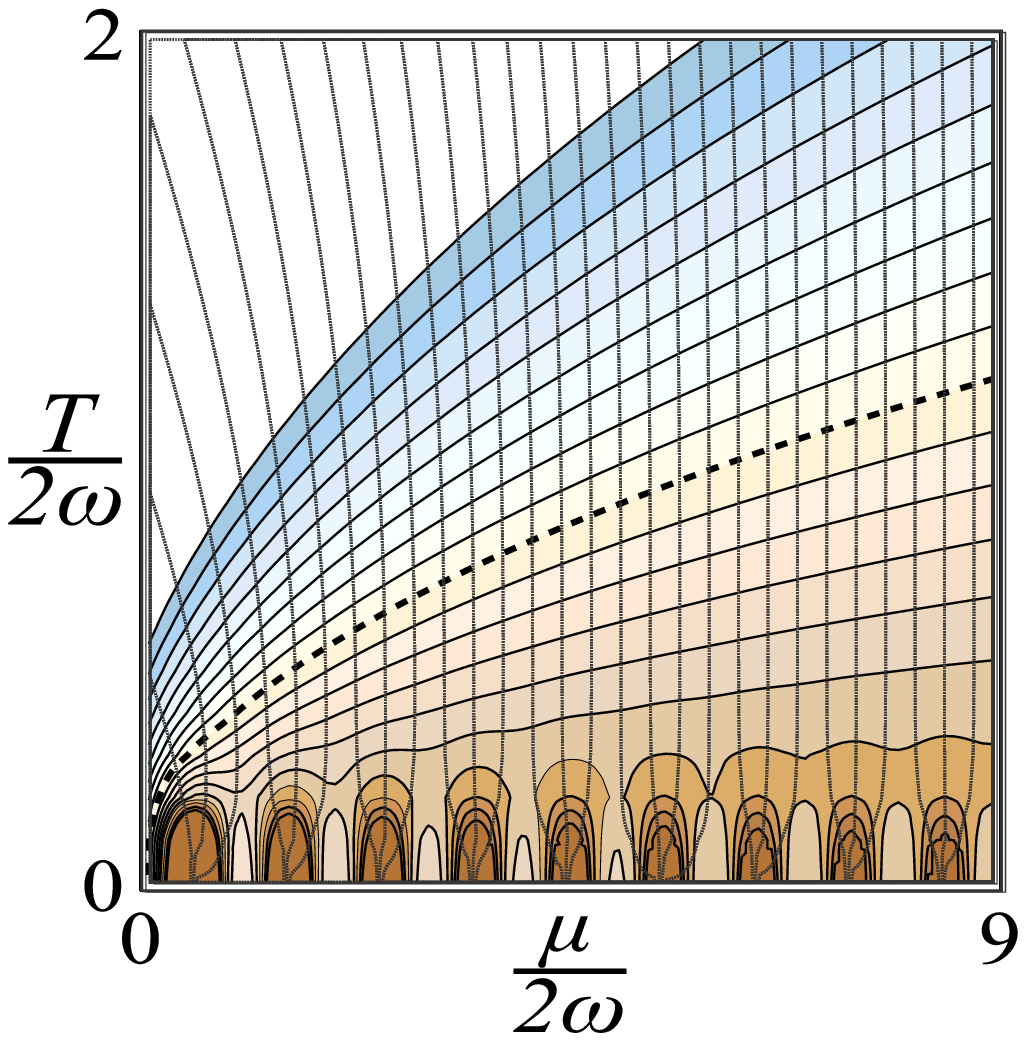}}
\subfigure[{}]{\includegraphics[height=1.4in]{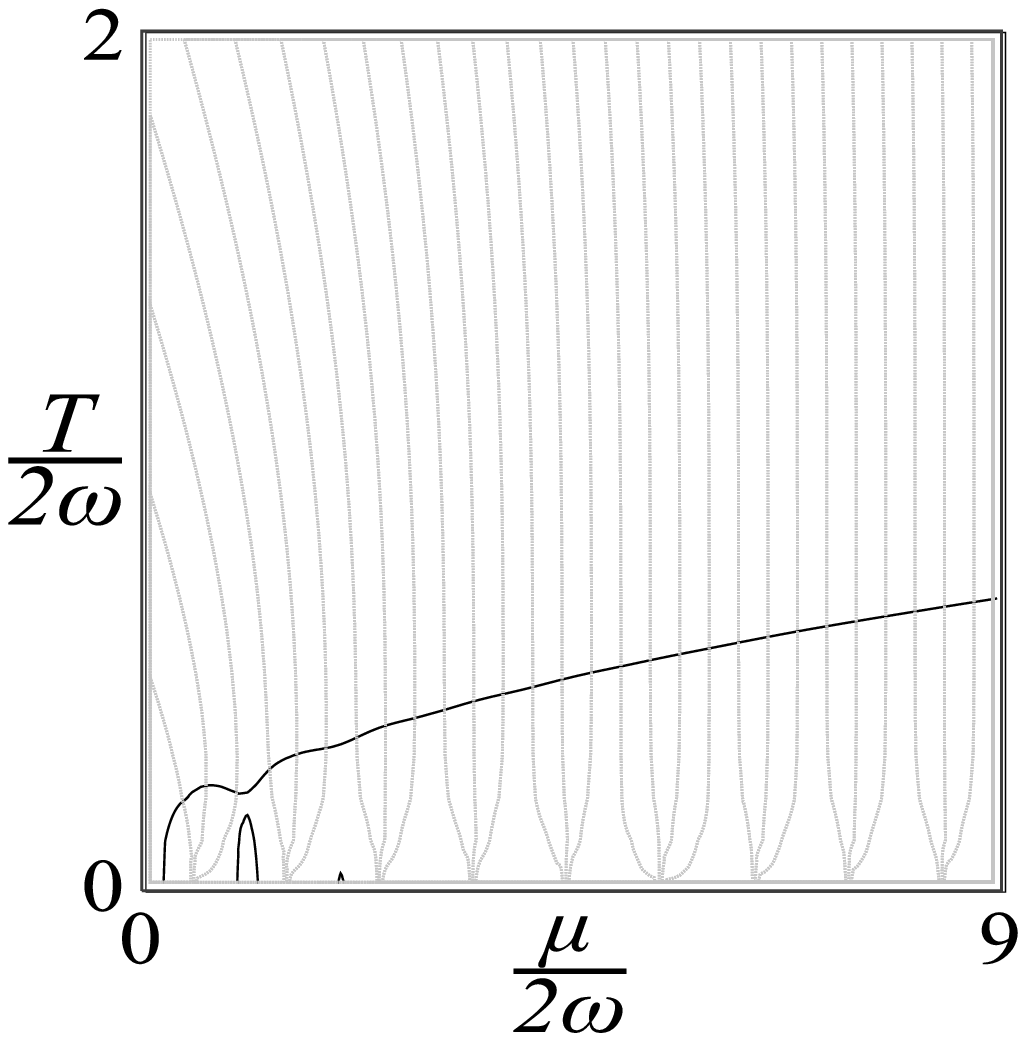}}
\subfigure[{}]{\includegraphics[height=1.4in]{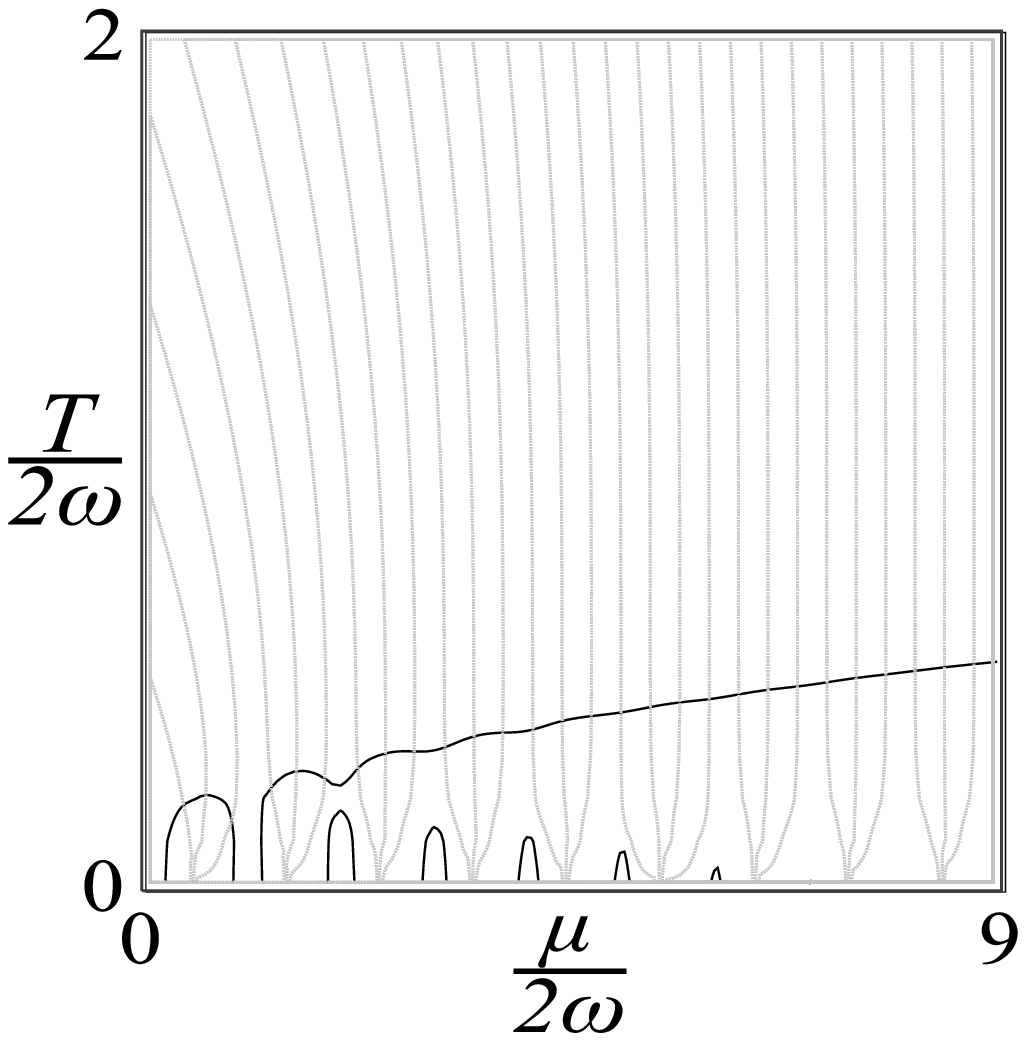}}
\subfigure[{}]{\includegraphics[height=1.4in]{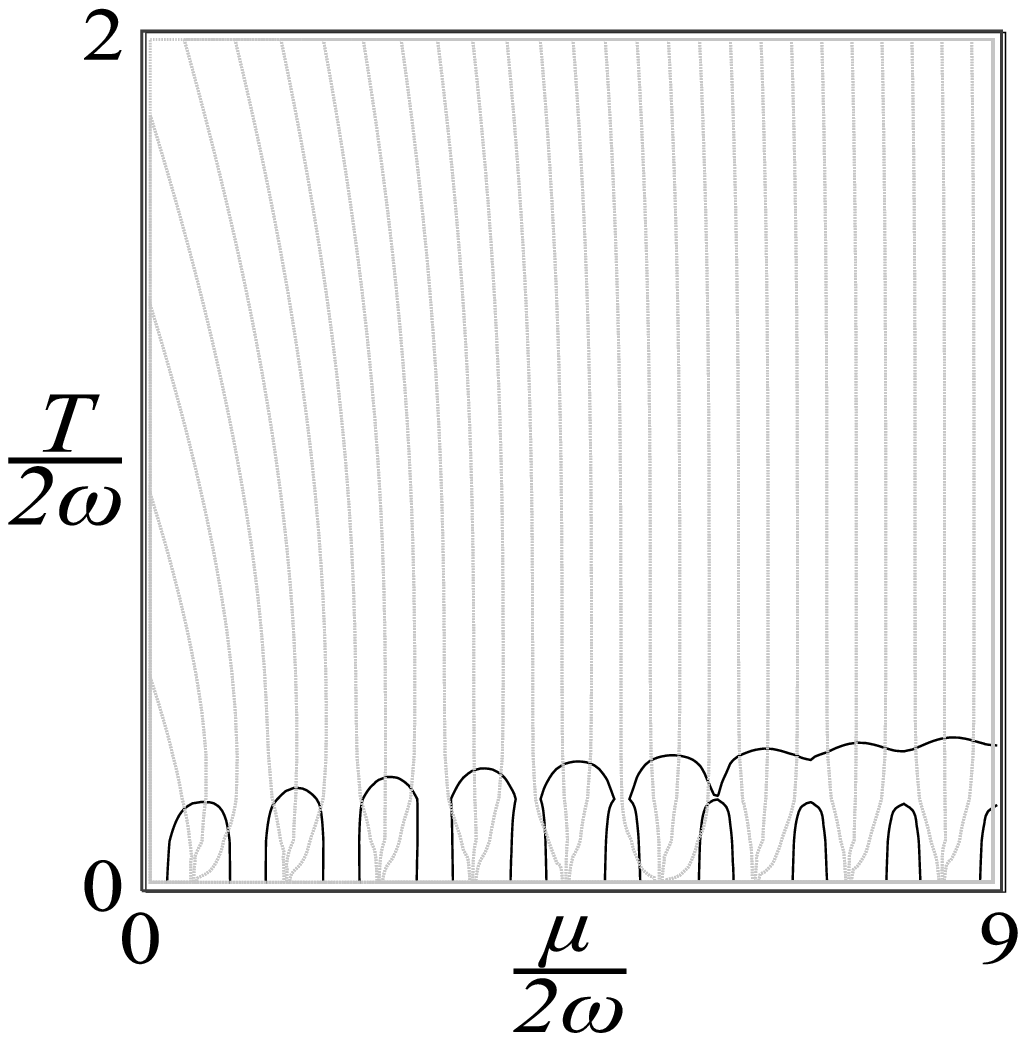}}
\caption{\label{Pairing}(color online) (a) Critical temperature as a function of chemical potential for several values of detuning $\widetilde{\nu}=\nu/(4m\omega a_z)$: the outermost curve is BEC limit $\widetilde{\nu}=-0.9$, the innermost is BCS limit $\widetilde{\nu}=+0.9$, unitarity $\nu=0$ is dashed ($\Delta\widetilde{\nu}=0.1$). The vertically stretching lines are normal state constant density contours, with increment $0.25 B/(2\pi)$. (b) through (d): The evolution of the transition line with detuning on the BCS side ($\widetilde{\nu}$ is $0.4$ in (b), $0.5$ in (c), $0.6$ in (d).}
\end{figure}

Figure \ref{Pairing} shows pairing second-order phase transitions in the limit $N\to\infty$. The deeper one goes into the BEC regime, the faster the growth of critical temperature with chemical potential (density). In the BCS regime, however, the paired phase breaks up into dome-shaped islands at low densities, obtained when chemical potential crosses a fermionic Landau level. Similar phenomena have been proposed in cuprates \cite{Zlatko3}. Once a paired state becomes a connected region at sufficiently large densities, there are trailing islands of unpaired states sitting between the Landau levels. Therefore, unpaired normal states can exist at $T=0$.

\begin{figure}[!]
\includegraphics[height=0.8in]{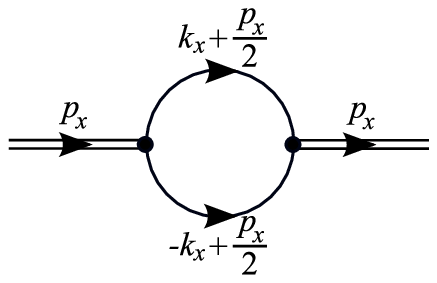}
\hspace{0.2in}
\raisebox{0.15in}{\includegraphics[height=0.7in]{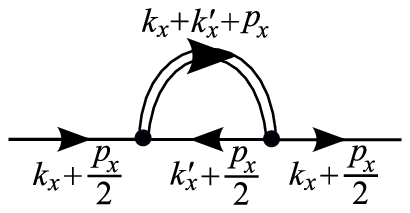}}
\caption{\label{SEexample} Feynman diagram labeling examples.}
\end{figure}

We now establish a crucial fact: ~(\ref{Pi1}) is independent of $p_x$ to all orders of $1/N$. First, the written lowest order term in the $1/N$ expansion involves fermion energies that do not depend on $\pm k_x + p_x/2$, and vertices ~(\ref{Vertex1}) that do not depend on $p_x$. This term defines the bare boson propagator $G_\Phi \propto N^{-1}$, using which we generate higher order corrections of ~(\ref{Pi1}). Let us apply the following labeling rules: each fermion propagator shall carry momentum $k_x+p_x/2$ in the arrow direction, while each boson propagator shall carry $q_x+p_x$ (see Fig.\ref{SEexample}). The transfer of $p_x$ is automatically conserved. Each added vertex takes momenta $k_{x1}+p_x/2$ and $k_{x2}+p_x/2$ at its fermionic terminals, but there is no dependence on $p_x$ since only the difference of momenta at the fermion terminals matters in ~(\ref{Vertex1}). The added bare boson lines take $p_x$ contributions, but according to ~(\ref{Pi1}) they do not depend on the transferred momentum. Therefore, order by order, no dependence on $p_x$ is introduced in any Feynman diagram.

This means that the poles of the exact boson Green's function are macroscopically degenerate in the normal phase, having no dependence on $p_x$. Pairing instability involves an infinite number of bosonic modes going soft at the same time. What are the physical consequences?

\begin{figure}[!]
\includegraphics[width=2.4in]{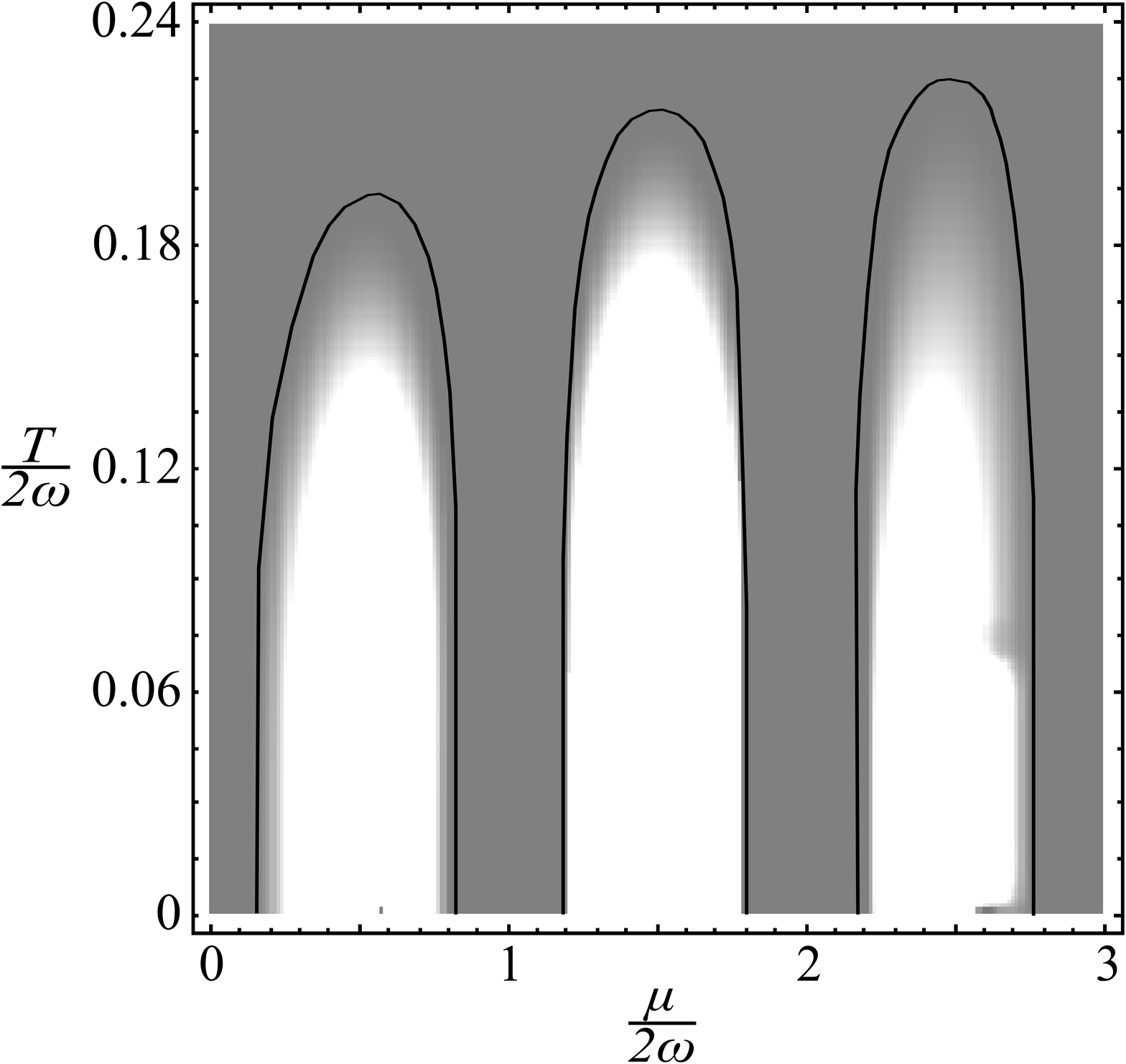}
\caption{\label{ALcomp}Density plot of $\Delta\mathcal{F} = \mathcal{F}_{\Tr{FL}} - \mathcal{F}_{\Tr{AL}}$, where $\mathcal{F}$ are free energy density minimums of the Abrikosov lattice (AL) and a flux lattice (FL) with $n_x=1$, $n_y=2$, $\delta q = \sqrt{B}$ ($\nu/(4m\omega a_z)=0.6$, $T=0$, $N=\infty$, the lowest bosonic Landau level only). $\Delta\mathcal{F}>0$ inside the paired regions bounded by the contours, but the halos around the bright domes have $\Delta\mathcal{F}/(2N\omega)<0.0005$.}
\end{figure}

Paired states are described by order parameters $\Phi(\rv)$ corresponding to static arrangements of vortices whose total number is $\mathcal{A}B/\pi$, $\mathcal{A}$ being the system area. In general, free energy density $\mathcal{F}$ is minimized by an order parameter with a periodic array of vortices:
\begin{equation}\label{OP2}
\Phi(\rv) = B^{-\frac{1}{4}} \sum_n \sum_{l=0}^{n_y-1}
  \phi_{n,l} \sum_j \Phi_{n, \delta q l + n_y \delta q j}(\rv) \ ,
\end{equation}
where the integer $n_y$ and momentum scale $\delta q$ determine the flux lattice periods $\Delta x = 2\pi / \delta q$ and $\Delta y = n_y \delta q (2B)^{-1}$ (note that $\Phi(\rv)$ has aperiodic gauge-dependent phase). The amplitudes $\phi_{n,l}$, as well as $\delta q$ and $n_y$, are obtained by minimizing $\mathcal{F}$, and in normal circumstances correspond to the triangular Abrikosov lattice: $\delta q = (2\pi\sqrt{3}B)^{1/2}$, $n_y=2$, $\phi_{0,1} = i\phi_{0,0}$. The minimum of $\mathcal{F}$ is well defined and unique deep inside the paired phase. However, near the pairing transition a large number of different paired states become rapidly competitive by free energy because: 1) the evolution of $\mathcal{F}$ with action parameters is smooth and 2) $\mathcal{F} \lbrack\Phi(\rv)\rbrack$ approaches $\mathcal{F}\lbrack 0 \rbrack$ at the same rate for all $\Phi(\rv)$ owing to the degeneracy of $\hat{\Pi} = \dd^2 \mathcal{F} / \dd \Phi(\rv)^2$. As an illustration we compare the free energy minimums of two flux lattices in Fig.\ref{ALcomp}.

It is now easy to see what role can be played by the relevant perturbations $u_n$ in ~(\ref{RotAct2}). These finite-range interactions do not need to be very strong to introduce sufficient mixing between competitive vortex states near the pairing transition, and melt the vortex lattice of the superfluid phase. The obtained normal states are strongly correlated vortex liquids. Their properties are not universal in the sense that the interactions $u_n$ that stabilize them reflect microscopic details of inter-atomic potential. Similarly, these perturbations control the properties of vortex cores in the superfluid. Unitarity looses many universal aspects, but paves the way for the emergence of stable vortex liquid states.

I am indebted to Olexei Motrunich, Gil Refael, Anton Burkov, Arun Paramekanti, Carlos Bolech and Satyan Bhongale for very helpful discussions. Numerical calculations were performed on Rice University supercomputers. This research was supported by Keck Fellowship.


\begin{thebibliography}{99}

% THEORY: Importance of phase fluctuations in superconductors with small superfluid density
\bibitem{Emery}
V.J.Emery, S.A.Kivelson;
Nature {\bf 374}, 434 (1995)

% EXPERIMENT: Nernst effect, N.P.Ong's group
\bibitem{Ong1}
Y.Wang, \etal;
Phys.Rev.B {\bf 64}, 224519 (2001)

% EXPERIMENT: Low temperature vortex liquid in \rm La_{2-x}Sr_xCuO_4
\bibitem{Ong2}
L.Li, \etal;
Nature Phys. {\bf 3}, 311 (2007)

% THEORY: Charge modulation, spin response, and dual Hofstadter butterfly in high-T-c cuprates
\bibitem{Zlatko1}
Z.Te\v{s}anovi\'{c};
Phys.Rev.Lett. {\bf 93}, 217004 (2004)

% THEORY: Putting competing orders in their place near the Mott transition
\bibitem{Subir1}
L.Balents, \etal;
Phys.Rev.B {\bf 71}, 144508 (2005)

% SURVEY OF STM EXPERIMENTS: Scanning tunneling spectroscopy of hightemperature superconductors
\bibitem{Fischer}
\O.Fischer, \etal;
Rev.Mod.Phys. {\bf 79}, 353 (2007)

% EXPERIMENT: Local ordering in the pseudogap state of BSCCO
\bibitem{Yazdani}
M.Vershinin, \etal;
Science {\bf 303}, 1995 (2004)

% EXPERIMENT: Imaging the two gaps of the high-tc superconductor Pb-BiSrCuO
\bibitem{Hudson}
M.C.Boyer, \etal;
Nature Phys. {\bf 3}, 802 (2007)

% EXPERIMENT: An Intrinsic Bond-Centered Electronic Glass with Unidirectional Domains in Underdoped Cuprates
\bibitem{Davis}
Y.Kohsaka, \etal;
Science {\bf 315}, 1380

% EXPERIMENT: The Ground State of the Pseudogap in Cuprate Superconductors
\bibitem{Valla}
T.Valla, \etal;
Science {\bf 314}, 1914

% EXPERIMENT: Many-Body Physics with Ultracold Gases
\bibitem{Bloch}
I.Bloch, J.Dalibard, W.Zwerger;
Rev.Mod.Phys. {\bf 80}, 885 (2008)

% THEORY: exact diagonalization, pairing in the lowest LL - phase separation, quantum Hall states (2/3, 1/2...)
\bibitem{Moller}
G.M\"{o}ller, Th.Jolicoeur, N.Regnault;
arXiv:0807.1034

% THEORY: Critical rotational frequency for superfluid fermionic gases accross Feshbach resonance
\bibitem{ZhaiHo}
H.Zhai, T.-L.Ho;
Phys.Rev.Lett. {\bf 97}, 180414 (2006)

% THEORY: Quantum Hall transition near a fermion Feshbach resonance in a rotating trap
\bibitem{YangZhai}
K.Yang, H.Zhai;
Phys.Rev.Lett. {\bf 100}, 030404 (2008)

% EXPERIMENT: fast rotation of an ultra-cold Bose gas (Rb)
\bibitem{Dalibard}
V.Bretin, \etal;
Phys.Rev.Lett. {\bf 92}, 050403 (2004)

% EXPERIMENT: long-lived vortex aggregates in rotating BEC
\bibitem{Engels}
P.Engels, \etal;
Phys.Rev.Lett. {\bf 90}, 170405 (2003)

%EXPERIMENT: BEC in fast rotation
\bibitem{Stock}
S.Stock, \etal;
Laser Phys.Lett. {\bf 2}, 275 (2005)

% THEORY: review of Feshbach resonances
\bibitem{Kohler}
T.K\"{o}hler, K.G\'{o}ral, P.S.Julienne;
Rev.Mod.Phys. {\bf 78}, 1311 (2006)

% THEORY: unitarity and population imbalance; Sp(N) large-N expansion
\bibitem{unitary}
P.Nikoli\'c and S.Sachdev;
Phys.Rev.A {\bf 75}, 033608 (2007)

% THEORY: Sp(N) large-N expansion
\bibitem{rvs}
M.Y.Veillette, D.E.Sheehy, L.Radzihovsky;
Phys.Rev.A {\bf 75}, 043614 (2007)

% THEORY: unitarity in optical lattices
\bibitem{optlat}
E.G.Moon, P.Nikoli\'c, S.Sachdev;
Phys.Rev.Lett. {\bf 99}, 230403 (2007)

% THEORY: Destruction by fluctuations of superconducting long-range order in the Abrikosov flux lattice
\bibitem{Moore}
M.A.Moore;
Phys.Rev.B {\bf 39}, 136 (1989)

% THEORY:
\bibitem{Zlatko2}
Z.Te\v{s}anovi\'{c};
Physica C {\bf 220}, 303 (1994)

% THEORY: Quantum melting and absence of Bose-Einstein condensation in 2D vortex matter
\bibitem{MacDonald}
J.Sinova, C.B.~Hanna, A.H.MacDonald;
Phys.Rev.Lett. {\bf 89}, 030403 (2002)

% THEORY: superconductivity in high Landau levels
\bibitem{Zlatko3}
M.Rasolt, Z.Te\v{s}anovi\'{c}; Rev.Mod.Phys. {\bf 64}, 709 (1992)


\end{thebibliography}
\end{document}